\documentclass[final,5p,times,twocolumn,number]{elsarticle}

\usepackage{amssymb}
\usepackage{lipsum}
\usepackage{amsmath}
\usepackage{braket}

\newcommand{\bfk}{{\bf k}_{\perp}}
\newcommand{\bfz}{{\bf z}_{\perp}}
\newcommand{\bfq}{{\bf q}_{\perp}}
\newcommand{\bfb}{{\bf b}_{\perp}} 
\newcommand{\bfa}{{\bf a}_{\perp}} 
\journal{Physics Letters B}

\begin{document}

\begin{frontmatter}



\title{Mechanical properties of proton in the momentum space}


\author[first]{Navpreet Kaur}
\affiliation[first]{organization={Department of Physics},
            addressline={Dr. B R Ambedkar National Institute of Technology Jalandhar}, 
            postcode={144008}, 
            state={Punjab},
            country={India}}
\ead[first]{knavpreet.hep@gmail.com}
\author[second]{Shubham Sharma}
\affiliation[second]{organization={Laboratory for Advanced Scientific Technologies of Mega-Science Facilities and
		Experiments},
	addressline={Moscow Institute of Physics and Technology (MIPT)}, 
	postcode={141700}, 
	state={Dolgoprudny},
	country={Russia}}
\ead[second]{s.sharma.hep@gmail.com}

\author[first]{Abi Jebarson A}
\ead[third]{abijebarson@gmail.com}

\author[first]{Harleen Dahiya}	
\ead[forth]{dahiyah@nitj.ac.in}
\begin{abstract}
We study the parametrization of the energy-momentum tensor for the case of a proton in momentum space in terms of gravitational transverse momentum-dependent distributions (TMDs). These gravitational TMDs are investigated with the inclusion of higher-twist contributions to predict the mechanical properties, specifically the transverse pressure and shear force distributions, along with the polarization-dependent $\Pi^q_S$ and $\Pi^q_A$ terms. The corresponding distributions are computed individually for both $u$ and $d$ quark flavors. The calculations have been performed in the light-cone framework using the spectator diquark model. A strong binding contribution to the transverse pressure is observed in the low-momentum space for both quark flavors of the proton.  
\end{abstract}




\end{frontmatter}




\section{Introduction}
\label{introduction}

At the partonic level, the dynamics of partons inside a proton can be described more completely by transverse momentum-dependent parton distributions (TMDs), as they provide a three-dimensional imaging of the complex internal structure of hadrons by considering the intrinsic transverse momentum $\bfk$ of partons with longitudinal momentum fraction $x$. This goes beyond the one-dimensional picture provided by standard parton distribution functions (PDFs) \cite{Diehl:2015uka, Pasquini:2008ax}.  Within quantum chromodynamics (QCD), semi-inclusive processes are described using TMD factorization \cite{Collins:2004nx, Polchinski:2002jw, Ji:2004wu} and encode a wealth of information regarding spin-momentum correlation through T-odd TMDs such as Sivers function \cite{Sivers:1989cc, Collins:2002kn} and Boer-Mulder function \cite{Boer:1997nt, Boer:2003cm}. These T-odd TMDs are calculated by considering final-state interaction (FSI) that involves the exchange of soft gluons between the active quark $q$ and the spectator \cite{Brodsky:2002cx, Brodsky:2003wu}. In the theoretical description of the semi-inclusive processes, the cross-sections are expanded in powers of $1/Q$. Here, $Q$ corresponds to the momentum transfer in the collision. A convolution of the twist-2 distributions involves the contribution of leading power of $1/Q$,  twist-3 distributions involve the contribution of subleading power of $1/Q$, and so on \cite{Qiu:1990xxa, Bacchetta:2006tn, Meissner:2007rx, Meissner:2009ww, Sharma:2023wha, Sharma:2022ylk}. The majority of work done so far is focused on twist-2 distributions \cite{Lyubovitskij:2020otz, Pasquini:2010af, Boffi:2009sh, Schweitzer:2013iva, AlizadehYazdi:2014dug, Schweitzer:2001sr, Tan:2022kgj}. Whereas higher-twist effects, which encode more intricate quark-gluon correlations, have received comparatively less attention \cite{Lu:2012gu, Zhu:2024awq, Pasquini:2018oyz, Lu:2015wja}.

While TMDs provide detailed information on the spin-momentum correlations of the hadron, they do not directly encode its internal mechanical properties. These properties are contained in the energy-momentum tensor (EMT), which governs the distribution of internal forces, such as the flux of inertia, transverse pressure, and shear force. Mechanical properties in spatial coordinates have been extensively studied employing different theoretical approaches, namely the pion mean-field approach grounded in the large Nc limit of QCD \cite{Won:2023zmf},  holographic model based on the AdS/CFT correspondence \cite{Nascimento:2025smf}, basis light-front quantization approach \cite{Nair:2024fit}, light-cone QCD sum rules \cite{Dehghan:2025ncw}, scalar diquark model \cite{Amor-Quiroz:2023rke} and bag model \cite{Lorce:2022cle}. The parametrization of the EMT in terms of gravitational TMDs has been discussed in Ref. \cite{Lorce:2023zzg} with an inclusion of leading and higher-twist distributions. In this work, we study this parametrization of TMD EMT for the case of proton and discuss the distributions of transverse pressure and shear force, along with the polarization-dependent $\Pi^q_S$ and $\Pi^q_A$ terms for both $u$ and $d$ quark flavors.

The light-cone framework has been adopted to study the relativistic quark dynamics of the proton. A spin-$\frac{1}{2}$, proton like particle is considered as a two-body system, comprising of an active quark that actively participates in the interaction with a real/virtual particle and a spectator diquark. The light-cone wave functions (LCWFs) are computed by considering the dipolar form factor of the proton-quark-diquark vertex. Depending upon the spin structure, a diquark can be scalar (spin-0) or axial-vector (spin-1). Further, these axial-vector diquarks are distinguished as isoscalar (ud-like) and isovector (uu-like) spectators \cite{Bacchetta:2008af}. For T-odd TMDs, there are two ways to include the FSI, either by modifying the light-cone wave functions (LCWFs) or by considering a kernel in the overlap form of LCWFs \cite{Brodsky:2002cx, Lu:2006kt, Pasquini:2019evu, Gamberg:2009uk}. We adopted the second method to determine the required leading and higher-twist T-odd TMDs for calculating mechanical properties in momentum space. 

The paper is presented as follows. The parametrization of TMD EMT and the description of the spectator diquark model is detailed in Sections \ref{Section2} and \ref{Section3}, respectively. In Section \ref{Section4}, we present the results for mechanical properties in the momentum space. Finally, we summarize the present work in the Section \ref{Section5}.
\section{Gravitational transverse momentum-dependent distributions \label{Section2}}
The local gauge-invariant EMT operator in QCD for the case of quark $q$ is 
\begin{eqnarray}
	T_q^{\alpha \beta}(r)=\bar{\Psi}(r)~ \gamma^\alpha \frac{i}{2} \overleftrightarrow{D}^\beta \Psi(r),
\end{eqnarray}
where $\overleftrightarrow{D}^\beta=\overrightarrow{\partial}^\beta-\overleftarrow{\partial}^\beta-2igA^\beta(r)$ \cite{Lorce:2023zzg}. A bilocal generalization is required to define the quark EMT at average momentum $k$. However, it is obscure due to the non-commutation of the covariant derivative with the Wilson line $\mathbb{W}$ \cite{Lorce:2012ce}. This indicates that $k$ cannot be identified with the kinetic momentum. In the light-cone gauge $A^+=0$, where the Wilson line reduces to unity, $k^\alpha=i \partial^\alpha$, i.e., the canonical momentum. Therefore, the appropriate object to generalize is the gauge-invariant canonical (gic) EMT, which is given by
\begin{eqnarray}
	T_{q,gic}^{\alpha \beta}(r)=\bar{\Psi}(r)~ \gamma^\alpha \frac{i}{2} \overleftrightarrow{D}^\beta_{pure} \Psi(r),
\end{eqnarray}
where $\overleftrightarrow{D}^\beta_{pure}$ is called pure-gauge covariant derivative \cite{Leader:2013jra, Lorce:2012rr, Chen:2008ag} and the quark bilocal gic EMT operator \cite{Ji:2003ak, Belitsky:2003nz} is written as
\begin{eqnarray}
	T_{q,gic}^{\alpha \beta}(r,k)  &=& k^\beta \int \frac{d^4z}{(2\pi)^4} e^{i k \cdot z} ~\bar{\Psi}\bigg(r-\frac{z}{2}\bigg)~ \gamma^\alpha~  \nonumber \\ &\times& \mathbb{W} \bigg(r-\frac{z}{2},r+\frac{z}{2}\bigg|n\bigg)~  \Psi\bigg(r+\frac{z}{2} \bigg).
\end{eqnarray}
The four-momentum $k$ represents the Fourier conjugate to the space-time distance $z$ between the quark operators $\bar \Psi$ and $\Psi$. It comes out to be unambigious due to the commutativity of the pure-gauge covariant derivatives  and the integration of $T^{\alpha \beta}_{q,gic} (r,k)$ over the four-momentum of quark gives $T_{q,gic}^{\alpha \beta}(r)$. The full unintegrated EMT in terms of the forward matrix elements of the operator $T^{\alpha \beta}_{q,gic} (r,k)$ is given by
\begin{eqnarray}
	\Theta^{\alpha \beta}_{q } (P,k,N,\lambda;\eta) = \frac{1}{2} \bra{P,\lambda} T^{\alpha \beta}_{q,gic} (0,k) \ket{P,\lambda}, \label{EMT}
\end{eqnarray}
where $P$ and $\lambda=(\Uparrow,\Downarrow$) denote the four-momentum and the initial state covariant spin vector, respectively. $N$ is the rescaling-invariant four-vector and the term $\eta=sign(n^0)$ with $n$ as the lightlike direction. Eq. (\ref{EMT}) follows the parity, hermiticity and time-reversal invariance. Considering integration over the light-cone energy of quark leads to the three-dimensional TMD EMT 
\begin{eqnarray}
	\mathbb{T}^{\alpha \beta}_q (P,x,\bfk,N,\lambda;\eta) &=& \int dk^- \Theta^{\alpha \beta}_q (P,k,N,\lambda;\eta) \nonumber \\ &=& \frac{1}{2} \int \frac{dz^- d^2 \bfz}{(2\pi)^3} e^{i k \cdot z}~i \partial^\beta_z \bra{P,\lambda} ~  \nonumber \\ &\times& \bar{\Psi}~\bigg(\frac{-z}{2}\bigg)~\gamma^\alpha ~ \mathbb{W} \bigg(\frac{-z}{2},\frac{z}{2}\bigg|n\bigg) ~\nonumber \\ &\times& \Psi\bigg(\frac{z}{2} \bigg) \ket{P,\lambda} ~\bigg|_{z^+=0}, 
\end{eqnarray}
in momentum space. The building blocks to construct the parametrization of TMD EMT are $g^{\alpha \beta}_T$, $\epsilon^{\alpha \beta}_T$, $P^\alpha$, $N^\alpha$ and $k^\alpha_T$. On considering all the independent rank-2 tensors, TMD EMT \cite{Lorce:2023zzg} is expressed as
\begin{eqnarray}
	\mathbb{T}^{\alpha \beta}_q &=& \frac{1}{P^+} \bigg[ P^\alpha P^\beta a_1 + N^\alpha N^\beta a_2 + \textbf{k}^\alpha_T \textbf{k}^\beta_T a_3 +P^\alpha N^\beta a_4 \nonumber \\ &+& N^\alpha P^\beta a_5 + P^\alpha \textbf{k}^\beta_T a_6 + \textbf{k}^\alpha_T P^\beta a_7 +N^\alpha \textbf{k}^\beta_T a_8 \nonumber \\ &+& \textbf{k}^\alpha_T N^\beta a_9 + M^2 g^{\alpha \beta}_T a_0 - \frac{\epsilon^{\textbf{k}_T \textbf{S}_T}}{M} \bigg\{ P^\alpha P^\beta a_{1T}^\perp + N^\alpha N^\beta a_{2T}^\perp \nonumber \\ &+& \textbf{k}^\alpha_T \textbf{k}^\beta_T a_{3T}^\perp + P^\alpha N^\beta a_{4T}^\perp + N^\alpha P^\beta a_{5T}^\perp + P^\alpha \textbf{k}^\beta_T a_{6T}^\perp \nonumber \\ &+& \textbf{k}^\alpha_T P^\beta a_{7T}^\perp + N^\alpha \textbf{k}^\beta_T a_{8T}^\perp + \textbf{k}^\alpha_T N^\beta a_{9T}^\perp + M^2 g^{\alpha \beta}_T a_{0T}^\perp \bigg\} \nonumber \\ &-& M \big\{ P^\alpha \epsilon^{\beta \textbf{S}_T} a_{1T} + P^\beta \epsilon^{\alpha \textbf{S}_T} a_{2T} + N^\alpha \epsilon^{\beta \textbf{S}_T} a_{3T} \nonumber \\ &+& N^\beta \epsilon^{\alpha \textbf{S}_T} a_{4T} + \textbf{k}^\alpha_T \epsilon^{\beta \textbf{S}_T} a_{5T} + \textbf{k}^\beta_T \epsilon^{\alpha \textbf{S}_T} a_{6T} \big\} \nonumber \\ &-
	& \lambda \big\{ P^\alpha \epsilon^{\beta \textbf{k}_T} a_{1L} + P^\beta \epsilon^{\alpha \textbf{k}_T} a_{2L} + N^\alpha \epsilon^{\beta \textbf{k}_T} a_{3L} \nonumber \\ &+& N^\beta \epsilon^{\alpha \textbf{k}_T} a_{4L} + \textbf{k}^\alpha_T \epsilon^{\beta \textbf{k}_T} a_{5L} + \textbf{k}^\beta_T \epsilon^{\alpha \textbf{k}_T} a_{6L} \big\} \bigg],
\end{eqnarray}
where $\textbf{k}^2_T=-\bfk^2$ in vector notation. The 22 coefficients $a_i$ are function of longitudinal momentum fraction $x$ and transverse momentum $\bfk$ (GeV) named as gravitational TMDs. $a_{0-9}$ corresponds to 10 polarization independent gravitational TMDs, $a_{(0-9)T}^\perp$ and $a_{(1-6)T}$ represent total 16 transverse polarization gravitational TMDs, and $a_{(1-6)L}$ are the 6 longitudinal polarized six gravitational TMDs. The quark correlator of usual TMDs \cite{Bacchetta:2006tn, Meissner:2009ww} is 
\begin{eqnarray}
	\Phi^{[\Gamma]}_q (P,x,\bfk,N,\lambda;\eta) &=& \int dk^- W^{[\Gamma]}_q (P,k,N,\lambda;\eta) \nonumber \\ &=& \frac{1}{2} \int \frac{dz^- d^2 \bfz}{(2\pi)^3} e^{i k \cdot z}~ \bra{P,\lambda}  \nonumber \\ &\times& \bar{\Psi}~\bigg(\frac{-z}{2}\bigg)~  \Gamma ~ \mathbb{W} \bigg(\frac{-z}{2},\frac{z}{2}\bigg|n\bigg)  \nonumber \\ &\times& \Psi\bigg(\frac{z}{2} \bigg) \ket{P,\lambda}~ \bigg|_{z^+=0}, \label{TMD}
\end{eqnarray}
where $\Gamma$ corresponds to the Dirac matrices \cite{Meissner:2009ww}. From Eqs. (\ref{EMT}) and (\ref{TMD}), we have the relation
\begin{eqnarray}
	 \mathbb{T}^{\alpha \beta}_q (P,x,\bfk,N,\lambda;\eta)&=&k^\beta \Phi^{[\gamma^\alpha]}_q (P,x,\bfk,N,\lambda;\eta). 
\end{eqnarray}
\begin{table*}[t]
	\centering
	\begin{tabular}{c c c c c} 
		\hline
		$~~~~\Gamma~~~~$ & $~~\Phi_{q \mathfrak{s}(\mathfrak{a})}^{[\Gamma]}(\Uparrow \Uparrow)+\Phi_{q \mathfrak{s}(\mathfrak{a})}^{[\Gamma]}(\Downarrow \Downarrow)~~$ & $~~\Phi_{q \mathfrak{s}(\mathfrak{a})}^{[\Gamma]}(\Uparrow \Uparrow)-\Phi_{q \mathfrak{s}(\mathfrak{a})}^{[\Gamma]}(\Downarrow \Downarrow)~~$ & $~~\Phi_{q \mathfrak{s}(\mathfrak{a})}^{[\Gamma]}(\Downarrow \Uparrow)+\Phi_{q \mathfrak{s}(\mathfrak{a})}^{[\Gamma]}(\Uparrow \Downarrow)~~$ & $~~\Phi_{q \mathfrak{s}(\mathfrak{a})}^{[\Gamma]}(\Downarrow \Uparrow)-\Phi_{q \mathfrak{s}(\mathfrak{a})}^{[\Gamma]}(\Uparrow \Downarrow)~~$ \\
		\hline
		\vspace{0.3em}
		$\gamma^+$ & 2$f_1^q$ & - & $\frac{-2 i k_y}{M} f_{1T}^{\perp(q)}$ & $\frac{2k_x}{M}f_{1T}^{\perp(q)}$ \\ 
		\vspace{0.3em}
		$\gamma^1_T$ & $\frac{2k_x}{P^+}f^{\perp(q)}$ & $\frac{-2 i k_y}{P^+} f_{L}^{\perp(q)}$  & $\frac{2 i k_x k_y}{M P^+} f_T^{\perp(q)}$ & $\frac{-2 M}{P^+} \bigg(f_T^{\prime(q)} + \frac{k^2_x}{M^2} f_T^{\perp(q)}\bigg)$ \\ 
		\vspace{0.3em}
		$\gamma^2_T$ & $\frac{2k_y}{P^+}f^{\perp(q)}$ & $\frac{2 i k_x}{P^+} f_{L}^{\perp(q)}$ & $\frac{2i M}{P^+} \bigg(f_T^{\prime(q)} + \frac{k^2_y}{M^2} f_T^{\perp(q)}\bigg)$ & $\frac{-2 k_x k_y}{M P^+} f_T^{\perp(q)}$ \\ 
		\vspace{0.3em}
		$\gamma^-$ & $\frac{2M^2}{(P^+)^2} f_3^q$ & - & $\frac{-2 i k_y M}{(P^+)^2} f_{3T}^{\perp(q)}$ & $\frac{2k_xM}{(P^+)^2}f_{3T}^{\perp(q)}$ \\ 
		\hline
	\end{tabular}
	\caption{Parametrization of the leading and higher-twist TMDs $\Phi^\Gamma_q$ as mentioned in Eqs. (\ref{Twist2}-\ref{Twist4}), for different combinations of $\lambda$ in initial and final state ($\Uparrow \Uparrow,\Uparrow \Downarrow, \Downarrow \Downarrow, \Downarrow \Uparrow$) \cite{Sharma:2024lal}.}
	\label{TableTMDs}
\end{table*}
The parametrization of TMD quark correlator for leading and higher-twist TMDs can be written as
\begin{eqnarray}
	\Phi^{[\gamma^+]}_q &=& f_1^q - \frac{\epsilon^{\textbf{k}_T \textbf{S}_T}_T}{M} f_{1T}^{\perp(q)}, \label{Twist2} \\
	\Phi^{[\gamma^i_T]}_q &=& \frac{M}{P^+}~ \bigg[\frac{\textbf{k}^i_T}{M} f^{\perp(q)}-\epsilon^{i\textbf{S}_T}_T f_T^{\prime(q)}-\lambda_l \frac{\epsilon^{i\textbf{k}_T}}{M} f_L^{\perp(q)} \nonumber \\ &-&\frac{\textbf{k}^i_T \textbf{k}^j_T-\frac{1}{2} \textbf{k}_T^2 g^{ij}_T}{M^2} \epsilon_{Tj\textbf{S}_T} f_T^{\perp(q)} \bigg], \label{Twist3} \\
	\Phi^{[\gamma^-]}_q &=& \bigg(\frac{M}{P^+}\bigg)^2 \bigg[f_3^q - \frac{\epsilon^{\textbf{k}_T \cdot \textbf{S}_T}_T}{M} f_{3T}^{\perp(q)} \bigg], \label{Twist4}
\end{eqnarray}
with $i=1,2$ and $\lambda_l$ as the longitudinal light-cone polarization. All the real-valued TMD coefficients are written in the compact notation and in general functions of $x$ and $\bfk$ (GeV). Out of eight TMDs, three $f_1^q$, $f^{\perp(q)}$ and $f_3^q$ are T-even TMDs and the remaining ones are T-odd TMDs. Table \ref*{TableTMDs} represents the parametrization of $\Phi^{[\gamma^+, \gamma^i_T, \gamma^-]}$ with respect to proton helicities $\lambda=$($\Uparrow, \Downarrow$) \cite{Sharma:2024lal}. Similar to the representation of 2D spatial distributions of pressure and shear force \cite{Lorce:2018egm}, the transverse pressure and shear force distributions in the momentum space can be computed from $\mathbb{T}^{\alpha \beta}_q$ as 
\begin{eqnarray}
	\mathbb{T}^{ij}_q &=& -g^{ij}_T \sigma^q +\bigg(\frac{1}{2} g^{ij}_T -\frac{\textbf{k}^i_T \textbf{k}^j_T}{\textbf{k}^2_T} \bigg) \Pi^q \nonumber \\ &+& \frac{\textbf{k}^i_T \epsilon^{j \textbf{k}_T} + \textbf{k}^j_T \epsilon^{i \textbf{k}_T}}{2\textbf{k}^2_T} \Pi_S^q + \epsilon^{ij} _T \Pi_A^q. \label{Tiq}
\end{eqnarray}
On replacing $\bfk$ with $\bfb$ by Fourier transformation, first two terms can also be found in the position space. The last two newly introduced transverse tensors are naively T-odd, provided that $\Pi^q_S$ and $\Pi^q_A$ are linear in the target polarization \cite{Lorce:2023zzg}. Considering the TMD EMT, we have
\begin{eqnarray}
	\mathbb{T}^{ij}_q &=& k^j_T ~ [\Phi^{\gamma^i_T}] \nonumber \\
	&=& \frac{1}{P^+} ~ \bigg[\biggl\{f^{\perp(q)} - \frac{M \epsilon^{\textbf{k}_T \textbf{S}_T}_T}{\textbf{k}^2_T}f^{-(q)}_T\biggr\} \nonumber \\
	&-& \biggl\{\lambda_l f_L^{\perp(q)} + \frac{M(\textbf{k}_T \cdot \textbf{S}_T)}{\textbf{k}^2_T} f^{+(q)}_T \biggr\}\bigg],
\end{eqnarray}
with 
\begin{eqnarray}
	2 \sigma^q &=&  -\frac{1}{P^+} \bigg[\textbf{k}^2_T ~f^{\perp(q)}-M \epsilon^{\textbf{k}_T \textbf{S}_T}_T f^{-(q)}_T\bigg], \label{sigma}\\
	2 \Pi_A^q &=&  -\frac{1}{P^+} \bigg[\lambda_l~ \textbf{k}^2_T ~f^{\perp(q)}_L+M (\textbf{k}_T \cdot \textbf{S}_T) f^{+(q)}_T\bigg], \label{Pi}
\end{eqnarray}
with $2 \sigma^q = \Pi^q$ and $2 \Pi_A^q=\Pi_S^q$ \cite{Lorce:2023zzg}. It is important to mention that these expressions include only twist-$3$ T-even and T-odd TMDs. The twist-$2$ and twist-$4$ distributions provide information of only longitudinal and transverse momentum densities, along with the transverse inertial flux. The expression of $f^{\pm(q)}_T$ TMD is
\begin{eqnarray}
	f^{\pm(q)}_T=f^{\prime(q)}_T\pm \frac{\bfk^2}{2M^2} f_T^{\perp(q)}.
\end{eqnarray}
\begin{table*}[t]
	\centering
	\begin{tabular}{l c c c} 
		\hline
		$\text{Diquark}$ & $m_{(\mathfrak{s}, \mathfrak{a}_u, \mathfrak{a}_d)}$ (GeV) & $\Lambda_{(\mathfrak{s}, \mathfrak{a}_u, \mathfrak{a}_d)}$ (GeV) & $c_{(\mathfrak{s}, \mathfrak{a}_u, \mathfrak{a}_d)}$ \\
		\hline
		\vspace{0.3em}
		$\text{Scalar}~(\mathfrak{s})$ & $0.822\pm0.053$ & $0.609\pm0.038$ & $0.847\pm0.111$ \\ 
		\vspace{0.3em}
		$\text{Axial-vector}~(\mathfrak{a}_u)$ & $1.492\pm0.173$  & $0.716\pm0.074$ & $1.061\pm0.085$ \\ 
		\vspace{0.3em}
		$\text{Axial-vector}~(\mathfrak{a}_d)$ & $0.890\pm0.008$ & $0.376\pm0.005$ & $0.880\pm0.008$ \\ 
		\hline
	\end{tabular}
	\caption{Free parameters used in the present calculation for quark mass $m_q=0.3$ GeV and proton mass $M=0.9$. GeV	}
	\label{TableParameters}
\end{table*}
\section{Diquark spectator model \label{Section3}}
Consider the general notation for four-vector notation as $a=[a^+,a^-,\bfa]$. For the target proton and an active quark with their respective masses (and helicites) $M$($\lambda$) and $m_q$($\lambda_q$), following are the four-momenta of proton and an active quark
\begin{eqnarray}
	P&=&\biggl[P^+,\frac{M^2}{2P^+},\textbf{0}_T \biggr], \\
	k&=&\biggl[xP^+,\frac{k^2+\bfk^2}{2xP^+},\bfk \biggr],
\end{eqnarray}
respectively. The spin-$\frac{1}{2}$ spinor for proton is denoted by $U(P,\lambda)$. The LCWFs for the scalar diquark $\mathfrak{s}$ with mass $m_\mathfrak{s}$ and spin-$\frac{1}{2}$ spinor $u(k,\lambda_q)$ is expressed as
\begin{eqnarray}
	\psi^{\lambda}_{\lambda_q} (x,\bfk) = \sqrt{\frac{k^+}{(P-k)^+}} \frac{1}{k^2-m^2_q} \bar{u} (k,\lambda_q) ~\mathcal{Y}_{\mathfrak{s}}~ U(P,\lambda),
	\label{ScalarWfn}
\end{eqnarray}
with  scalar vertex $\mathcal{Y}_{\mathfrak{s}}=ig_{\mathfrak{s}}(k^2) \bf{1}$. Isoscalar ($ud-$like) and isovector ($uu-$like) diquarks (for respective $u$ and $d$ as an active quarks) are denoted by $\mathfrak{a}=\mathfrak{a}_u$, $\mathfrak{a}_d$, respectively. The LCWFs for the axial-vector diquarks $\mathfrak{a}$ with mass $m_\mathfrak{a}$, four-momentum ($P-k$) and helicity $\lambda_\mathfrak{a}$ are expressed as
\begin{eqnarray}
	\psi^{\lambda}_{\lambda_q \lambda_{\mathfrak{a}}}  (x,\bfk) &=& \sqrt{\frac{k^+}{(P-k)^+}} \frac{1}{k^2-m^2_q} \nonumber \\ &\times& \bar{u} (k,\lambda_q) \epsilon^\ast_\mu (P-k,\lambda_{\mathfrak{a}}) \cdot \mathcal{Y}_{\mathfrak{a}}^\mu~ U(P,\lambda),
	\label{VectorWfn}
\end{eqnarray}
with vertex $\mathcal{Y}_{\mathfrak{a}}^\mu=ig_{\mathfrak{a}}(k^2) \gamma^\mu \gamma_5 /\sqrt{2}$ \cite{Jakob:1997wg, Gamberg:2007wm}. We consider a dipolar form factor for spectator diquarks and its form is
\begin{eqnarray}
	g_{\mathfrak{s}(\mathfrak{a})}(k^2)=	g_{\mathfrak{s}(\mathfrak{a})} \frac{p^2-m_q^2}{|p^2-\Lambda^2_{\mathfrak{s}(\mathfrak{a})}|}.
\end{eqnarray}
Here, the quantity $\Lambda_{\mathfrak{s}(\mathfrak{a})}$ is the cutoff parameter. The explicit form of these LCWFs can be found in Ref. \cite{Bacchetta:2008af}. Since $\lambda$ in the initial and final state can have possible combinations of ($\Uparrow \Uparrow,\Uparrow \Downarrow, \Downarrow \Downarrow, \Downarrow \Uparrow$), we distinguish between the initial and final state of both proton by introducing a prime notation. In a similar pattern, helicity of an active quark $\lambda_q$ can also have such four combinations. Henceforth, $\lambda$($\lambda_q$) denotes the helicity of the initial state, while $\lambda^\prime$($\lambda_q^\prime$) represents the final state helicity of proton (active quark). In order to evaluate the different components $\sigma^q$, $\Pi^q$, $\Pi^q_S$ and $\Pi^q_A$ of TMD EMT $\mathbb{T}^{ij}_q$ component, the  required TMD correlator $\Phi^{[\Gamma]}_{q \mathfrak{s}(\mathfrak{a})}$ (as expressed in Table \ref{TableTMDs}) in terms of overlap form of LCWFs for scalar and axial-vector diquarks can be written as
\begin{eqnarray}
	\Phi^{[\Gamma]}_{q \mathfrak{s}}(\lambda,\lambda^\prime) &=&\frac{1}{16 \pi^3} \sum_{\lambda_{q}} \sum_{\lambda_{q^\prime}} \psi^{\lambda^\prime}_{\lambda_{q^\prime}}(x,\bfk)~ \psi^{\lambda}_{\lambda_{q}}(x,\bfk) \nonumber \\ &\times& \frac{\bar{u}(k,\lambda^\prime) \Gamma u(k,\lambda) }{2xP^+}, \\
	\Phi^{[\Gamma]}_{q \mathfrak{a}}(\lambda,\lambda^\prime) &=&\frac{1}{16 \pi^3} \sum_{\lambda_{q}} \sum_{\lambda_{\mathfrak{a}}} \sum_{\lambda_{q^\prime}} \psi^{\lambda^\prime}_{\lambda_{q^\prime} \lambda_{\mathfrak{a}}}(x,\bfk)~  \nonumber \\ &\times&  \psi^{\lambda}_{\lambda_{q}\lambda_{\mathfrak{a}}}(x,\bfk) \frac{\bar{u}(k,\lambda^\prime) \Gamma u(k,\lambda) }{2xP^+},
\end{eqnarray}
respectively. The free parameters of the model are computed using $MINUIT$ program \cite{Bacchetta:2008af} by fitting the normalized $f_1^q$ TMD. New coupling constants $c_{(\mathfrak{s}, \mathfrak{a}_u, \mathfrak{a}_d)}$ have been introduced which differ from the original $g_{(\mathfrak{s}, \mathfrak{a}_u, \mathfrak{a}_d)}$ by normalization constant $N_{(\mathfrak{s}, \mathfrak{a}_u, \mathfrak{a}_d)}$. The $f_1^q$ TMD is normalized by
\begin{eqnarray}
	\int dx \int d^2 \bfk ~ f^q_1(x,\bfk) = 1,
\end{eqnarray}
and the contribution for $u$ and $d$ quark flavor is given by $f_1^u = c_s^2 ~ f^u_1 + c_{\mathfrak{a}_u}^2 ~ f^u_1$ and $f_1^d = c_{\mathfrak{a}_d}^2 ~ f^d_1$. For the sake of completeness, all the free parameters used in the present calculations are mentioned in the Table \ref{TableParameters}. 

To generate non-zero T-odd TMDs, a perturbative Abelian gluon exchange approximation of the gauge link, used in Refs. \cite{Brodsky:2002cx, Brodsky:2002rv, Gurjar:2022rcl}, has been extended to the higher-twist TMDs and the kernel to include this exchange is given by 
\begin{eqnarray}
	i G(x,\bfq)=\frac{\alpha_s C_f}{\pi \bfq^2},
\end{eqnarray}
where $\alpha_s=0.3$ and $C_f=\frac{4}{3}$ are coupling constants and the color factor, respectively. The momentum $q_T$ denotes the momentum carried by the gluon, which is given by $\bfq=\bfk-\bfk^\prime$. For T-odd TMDs, overlap form of LCWFs for scalar and axial-vector diquarks are modified as
\begin{eqnarray}
	\Phi^{[\Gamma]}_{q \mathfrak{s}}(\lambda,\lambda^\prime) &=& i \int \frac{ d^2 \bfq}{16 \pi^3} \sum_{\lambda_{q}} \sum_{\lambda_{q^\prime}} \psi^{\lambda^\prime}_{\lambda_{q^\prime}}(x,\bfk)G(x,\bfq)  \nonumber \\ &\times& \psi^{\lambda}_{\lambda_{q}}(x,\bfk^\prime)~\frac{\bar{u}(k,\lambda^\prime) \Gamma u(k^\prime,\lambda) }{2xP^+}, \\
	\Phi^{[\Gamma]}_{q \mathfrak{a}}(\lambda,\lambda^\prime) &=& i \int \frac{d^2\bfq}{16 \pi^3} \sum_{\lambda_{q}} \sum_{\lambda_{\mathfrak{a}}} \sum_{\lambda_{q^\prime}} \psi^{\lambda^\prime}_{\lambda_{q^\prime} \lambda_{\mathfrak{a}}}(x,\bfk)~   \nonumber \\ &\times&  G(x,\bfq)\psi^{\lambda}_{\lambda_{q}\lambda_{\mathfrak{a}}}(x,\bfk^\prime) ~\frac{\bar{u}(k,\lambda^\prime) \Gamma u(k,\lambda) }{2xP^+},
\end{eqnarray}
respectively.

\section{Discussion \label{Section4}}
\begin{figure}
	\centering 
	(a)\includegraphics[width=7.0cm]{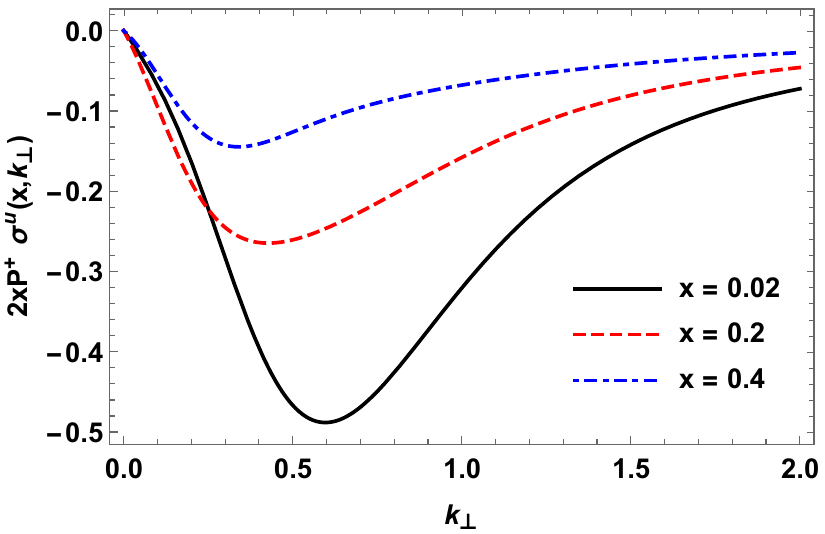}	
	\hspace{0.03cm}
	(b)\includegraphics[width=7.0cm]{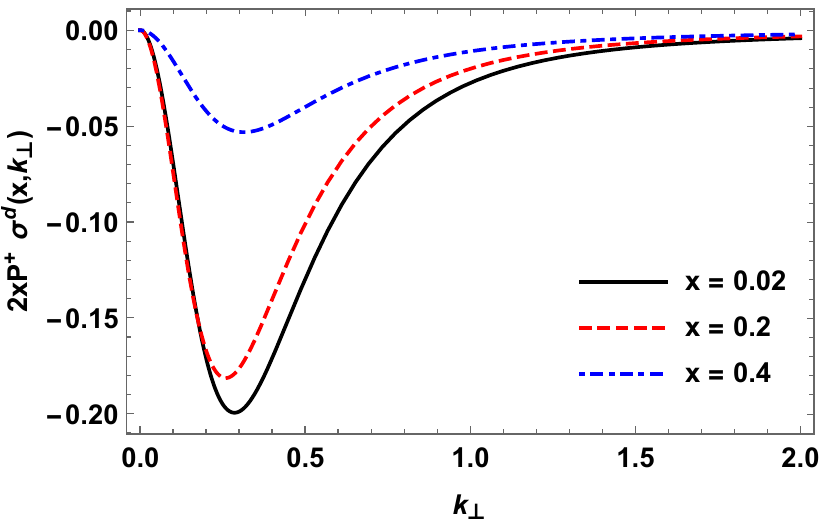}
	\caption{The transverse pressure distribution $\sigma^q$ of (a) $u$ and (b) $d$ quark flavors of proton as a function of transverse momentum $\bfk$ (GeV) at fixed values of longitudinal momentum fraction $x$.} 
	\label{fig_1}
\end{figure}

Fig. \ref{fig_1} represents the transverse pressure distribution $\sigma^q$ for (a) $u$ and (b) $d$ quark flavors of proton, expressed in Eq. (\ref{sigma}), as a function of intrinsic transverse momentum $\bfk$ (GeV) for fixed values of longitudinal momentum fraction $x$. The second term of the expression $\sigma^q$ contains $\epsilon^{\textbf{k}_T \textbf{S}_T}_T$ and its maximum contribution is considered when $\textbf{k}_T$ and $\textbf{S}_T$ are perpendicular to each other. For the $u$ quark flavor, the distribution has a peak at smaller values of $\bfk$ (GeV) in the negative region. As $\bfk$ (GeV) increases, the magnitude of the distribution gradually decreases. With an increase in the value of $x$, the magnitude of transverse pressure goes down with the shifting of a peak towards smaller values of $\bfk$ (GeV). It indicates a strong attractive pressure contribution for the low-$\bfk$ region. Compared to the $u$ quark flavor, the pressure distribution of the $d$ quark flavor exhibits a smaller magnitude of peak at smaller $\bfk$ (GeV) and falls off more rapidly with increasing $\bfk$ (GeV). Thus, both quark flavors indicate strong binding behavior only in the low-$\bfk$ region.

\begin{figure}
	\centering 
	(a)\includegraphics[width=7.0cm]{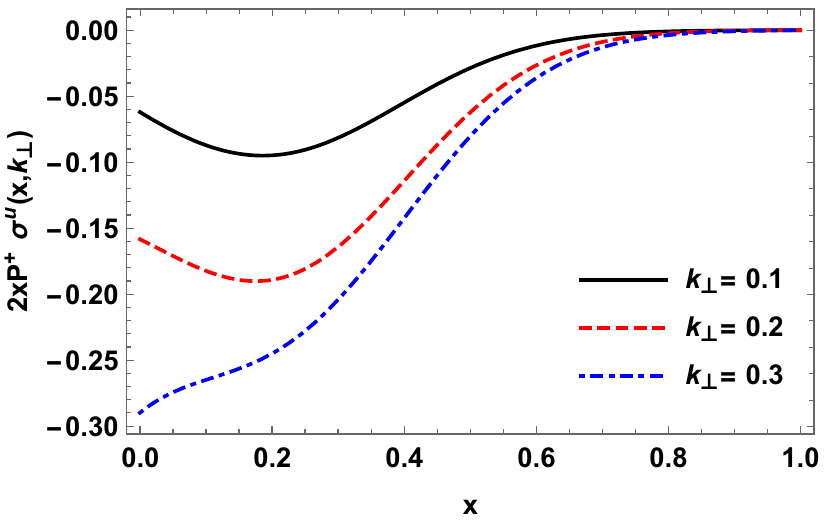}	
	\hspace{0.03cm}
	(b)\includegraphics[width=7.0cm]{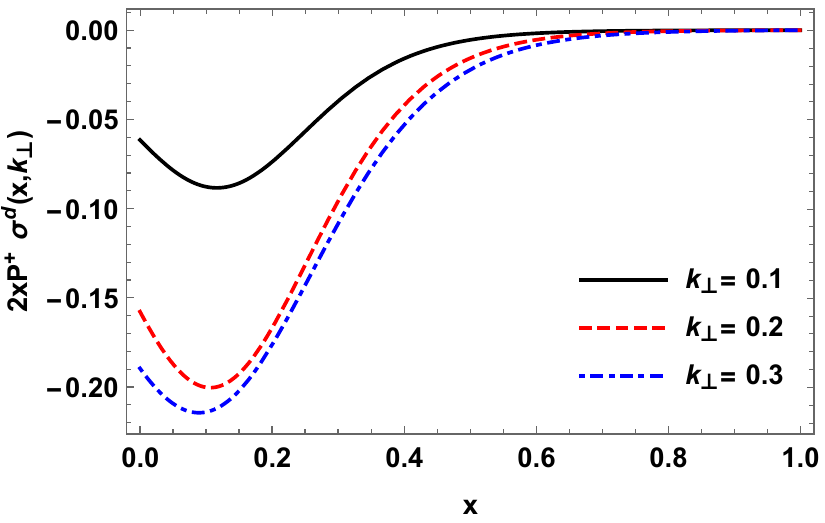}
	\caption{The transverse pressure distribution $\sigma^q$ of (a) $u$ and (b) $d$ quark flavors of proton as a function of longitudinal momentum fraction $x$ at fixed values of transverse momentum $\bfk$ (GeV).} 
	\label{fig_2}
\end{figure}

As both quark flavors show strong confining pressure contribution in small $\bfk$ region, in Fig. \ref{fig_2}, the distributions of transverse pressure $\sigma^q$ for (a) $u$ and (b) $d$ quark flavors of proton as a function of longitudinal momentum fraction $x$ for fixed small values of intrinsic transverse momentum $\bfk$ (GeV) are presented to observe the distribution on the entire region of $x$. We find that the $u$ quark flavor exhibits an attractive pressure contribution across all values of the longitudinal momentum fraction $x$ and vanishes for $x\rightarrow 1$. Relative to the $u$ quark flavor, the transverse pressure distribution of the $d$ quark saturates faster and vanishes earlier as a function of $x$. Hence, irrespective of the transverse momentum carried by the quarks, the pressure distribution becomes independent of longitudinal momentum fraction when $x\rightarrow1$. As per the Eq. (\ref{sigma}), the shear force is just half of the pressure. So, it has similar qualitative behavior. 

\begin{figure}
	\centering 
	(a)\includegraphics[width=7.0cm]{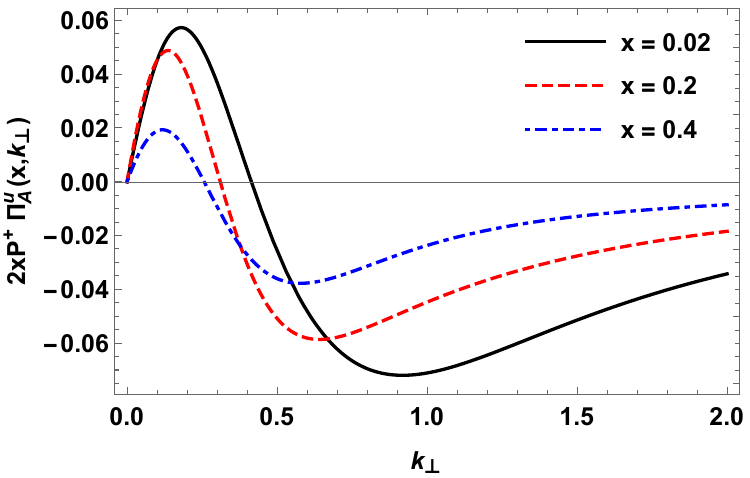}	
	\hspace{0.03cm}
	(b)\includegraphics[width=7.0cm]{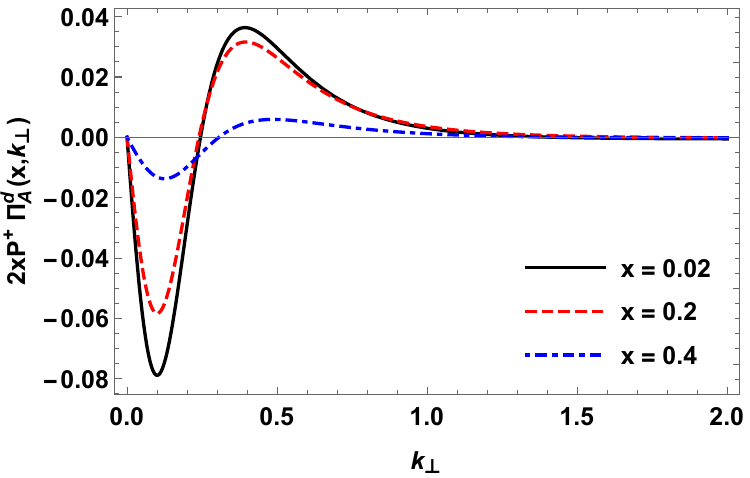}
	\caption{The distribution of $\Pi^q_A$ of (a) $u$ and (b) $d$ quark flavors of proton as a function of transverse momentum $\bfk$ (GeV) at fixed values of longitudinal momentum fraction $x$.} 
	\label{fig_3}
\end{figure}

The other two quantities introduced in Eq. (\ref{Tiq}) are $\Pi^q_A$ and $\Pi^q_S$, related as $2 \Pi_A^q=\Pi_S^q$, contain the contribution of only T-odd twist-3 distributions. Fig. \ref{fig_3} represents the term $\Pi^q_A$ for (a) $u$ and (b) $d$ quark flavors of proton as a function of intrinsic transverse momentum $\bfk$ (GeV) for fixed values of longitudinal momentum fraction $x$. The distribution for the $u$ quark flavor as an increasing function of $\bfk$ (GeV) is found to have a positive distribution in the low-$\bfk$ region, crosses zero, attains a negative peak, and then the magnitude falls. With an increase in the value of $x$, the magnitude of the distribution decreases, and the nodal point shifts toward smaller values of $\bfk$ (GeV). For the $d$ quark flavor, the sign of the distribution changes compared to the $u$ quark flavor. A negative peak has been observed for extremely small values of $\bfk$ (GeV), crossed zero, attained a positive peak, and then fell off. As an increasing function of $x$, the magnitude of the distribution gradually diminishes. In the high-$\bfk$ region, the distribution vanishes irrespective of the longitudinal momentum fraction carried by quarks.

\begin{figure}
	\centering 
	(a)\includegraphics[width=7.0cm]{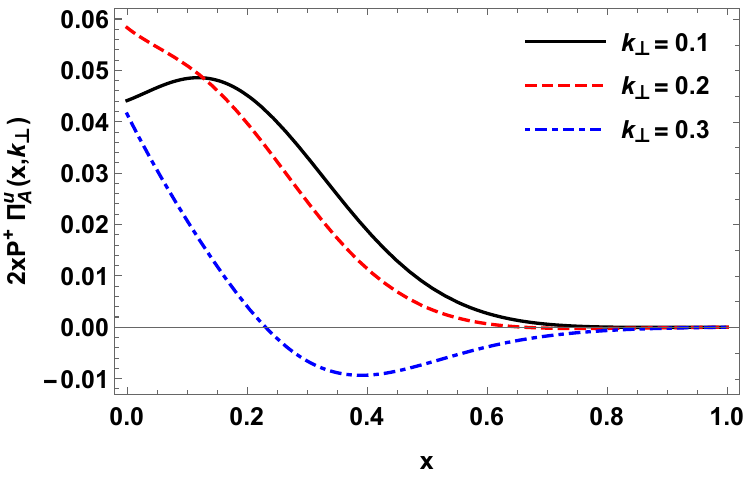}	
	\hspace{0.03cm}
	(b)\includegraphics[width=7.0cm]{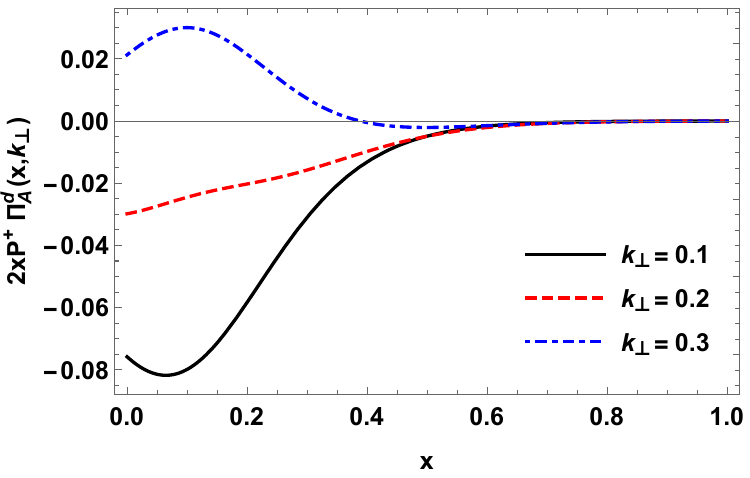}
	\caption{The distribution of $\Pi^q_A$ of (a) $u$ and (b) $d$ quark flavors of proton as a function of longitudinal momentum fraction $x$ at fixed values of transverse momentum $\bfk$ (GeV).} 
	\label{fig_4}
\end{figure}

Similarly, to observe the distribution of $\Pi^q_A$ on the entire range of longitudinal momentum fraction $x$, it is plotted in Fig. \ref{fig_4} at fixed values of intrinsic transverse momentum $\bfk$ (GeV) for (a) $u$ and (b) $d$ quark flavors of proton. $\Pi^q_A$ is found to have a positive distribution for $\bfk=0.1$ GeV and $\bfk=0.2$ GeV, whereas at  $\bfk=0.3$ GeV, the distribution contains a node as the distribution becomes negative for the intermediate region of $x$. For the case of $d$ quark flavor, the quantity $\Pi^q_A$ has negative distributions for $\bfk=0.1$ GeV and $\bfk=0.2$ GeV. At $\bfk=0.3$ (GeV), the distribution shows a node with a positive distribution in the low-$x$ region. As a function of $x$, it becomes slightly negative for an intermediate region and then vanishes. We find that similar to the transverse pressure distribution, the quantity $\Pi^q_A$ also becomes independent of longitudinal momentum fraction in the high $x$ region, irrespective of transverse momentum carried by the quarks.
\section{Summary and Conclusions \label{Section5}}
In this letter, we investigate the quark ($u,d$) contribution to the proton’s QCD energy-momentum tensor (EMT) in momentum space via gravitational transverse momentum-dependent distributions (TMDs). These are parametrized in terms of the usual T-even and T-odd TMDs to study the mechanical properties. In addition to the transverse pressure $\sigma^q$ and shear force $\Pi^q$, we analyze two tensor objects, $\Pi^q_A$ and $\Pi^q_S$, as a function of longitudinal momentum fraction $x$ and transverse momentum $\bfk$ (GeV). The required leading and higher-twist TMDs have been computed by using the spectator diquark model in the light-cone framework. 
We find that the transverse pressure exhibits a confining behavior for both flavors, with a stronger and more extended contribution from the $u$ quark than the $d$ quark. The distribution $\Pi_A^q$ shows opposite signs for $u$ and $d$ quarks, and a nodal structure for the $u$ flavor, reflecting complementary behavior between low- and high-$\mathbf{k}_\perp$ regions. Although momentum-space EMT components are not directly measurable, their relation to TMDs provides a valuable framework to explore hadron mechanical properties, including the role of higher-twist effects.

%

\section*{Acknowledgements}
H.D. would like to thank  the Science and Engineering Research Board, Anusandhan-National Research Foundation, Government of India under the scheme SERB-POWER Fellowship (Ref No. SPF/2023/000116) for financial support.



\end{document}